\documentstyle[aps,twocolumn,epsfig]{revtex}

\begin{document}

\input epsf

\title{Non-conservation of Density of States in Bi$_2$Sr$_2$CaCu$_2$O$_y$:
Coexistence of Pseudogap and Superconducting gap }

\author{Anjan K. Gupta\thanks{Present Address: CNRS-CRTBT, 25 Ave des 
Martyres, BP166, 38042, Grenoble, C\'{e}dex9, France} and K. -W. Ng}

\address{Department of Physics and Astronomy, University of Kentucky,
Lxington,  Kentucky, 40506-0055}

\maketitle

\begin{abstract}
Abstract: The tunneling spectra obtained within the ab-plane of 
Bi$_2$Sr$_2$CaCu$_2$O$_y$
(Bi2212) for temperatures below and above the critical temperature (T$_c$) 
are analyzed.
We find that the tunneling conductance spectra for the underdoped compound 
in the
superconducting state do not follow the conservation of states rule. There 
is a consistent
loss of states for the underdoped BI2212 implying an underlying depression 
in the density
of states (DOS) and hence the pseudogap near the Fermi energy (E$_F$). Such 
an underlying
depression can also explain the peak-dip-hump structure observed in the 
spectra.
Furthermore, the conservation of states is recovered and the dip-hump 
structure disappears
after normalizing the low temperature spectra with that of the normal state. 
We argue that
this is a direct evidence for the coexistence of a pseudogap with the 
superconducting gap.

\end{abstract}

\vspace{0.2cm}
The existence of a pseudogap in high-T$_c$ cuprate superconductors is a very 
well
established fact\cite{pg-review}. However, there is no agreement among 
theorists
regarding its origin. One popular belief is the pre-formation of pairs well 
above T$_c$
without any coherence\cite{emery-preformation}. Such pre-formation should 
carry some
signatures of the existence of pairs above T$_c$ such as a diamagnetic 
response or
Andreev reflection. No such supporting evidence in this direction has come 
so far, partly
because of the experimental difficulties. Independent of such a lack of 
direct evidence for
pre-formed pairs, there is ample evidence in support of a loss of states 
near
E$_F$\cite{pg-review}. This loss of states can also arise because of other 
possibilities,
for instance, a lattice distortion or a magnetic instability as suggested by 
various
authors\cite{pg-bandgap}.

The superconducting energy gap (2$\Delta$) of these high-T$_c$ 
superconductors
is a monotonic function of doping, with the gap decreasing as the carrier
concentration increases (while T$_c$ is peaked at a doping called optimum
doping)\cite{pg-review}. The pseudogap is known to exist in the underdoped 
and
slightly overdoped regime. If the pseudogap is different from the
superconducting gap, the former will continue to exist as a depression in 
the
DOS together with the latter as the temperature is reduced below T$_c$. 
Recent
intrinsic tunneling measurements on mesa structures have shown the existence
of a pseudogap together with the superconducting gap below 
T$_c$\cite{josephson}. In
these experiments, the superconducting gap is found to vanish above T$_c$ or 
in high
magnetic fields with the pseudogap remaining\cite{josephson-mag}. Further, a
peak-dip-hump structure in the DOS below T$_c$ seems to arise as a result of 
the two
coexisting gaps.

Such a peak-dip-hump structure has also been observed earlier in the c-axis 
as well as
ab-plane tunneling spectra at low
temperatures\cite{pg-renner,miyakawa-john,matsuda,gupta-rapcom}. 
Furthermore, the
normalized quasiparticle peak height at low temperatures is found to be 
higher for higher
doping\cite{pg-renner,miyakawa-john,matsuda}, without much change in the 
zero
bias conductance. This fact implies that, even in the superconducting state,
more states are lost near (E$_F$) with underdoping. Some other measurements,
like specific heat\cite{loram-sp-heat} and NMR \cite{tallon-nmr}, have also
concluded this kind of loss of states and hence the coexistence of the two 
gaps.
A direct consequence of this coexistence is the violation of conservation of 
states rule.
As a result of this, there will be a loss of states as we go from overdoped 
to underdoped
regime. We believe that the reduction in the normalized peak height with 
underdoping is a
consequence of this depression in the DOS and the peak-dip-hump structure is 
a result of
the superposition of the two gaps, one being slightly larger than the other.

The total number of states must be conserved within a band. So a depression
at E$_F$ means that some of the states are pushed out from near E$_F$ to
higher or lower energies (above or below E$_F$).  For example, when
superconductivity sets in at T$_c$, the states within the gap region,
$\pm\Delta$, are pushed out to higher energies; a majority of these states 
are
concentrated at $\pm\Delta$ as singularities. The average DOS in between 
$\pm
\epsilon$ ($\epsilon>\Delta$) stays the same as the background DOS (i.e. DOS
at $|eV|>\Delta$). Hence, in the tunneling experiments, if the average
tunneling conductance is less than the background conductance, there is an
underlying depression in the DOS. Our interpretation of the pseudogap as a
depression in the DOS at E$_F$ differs from the ARPES results, where the
pseudogap is interpreted as a true energy gap from the leading edge
analysis of the photoemission spectra\cite{arpes}. This kind of
state-conservation analysis is difficult to carry out for the photoemission
measurements, partly, because of a poor understanding of the photoemission
line-shape, and also because the data are limited to below E$_F$.

For a conventional BCS superconductor the DOS below T$_c$ is given by,
$N_{sup}(E)=N(E)|E|/\surd(E^2-\Delta^2)$. If we ignore the thermal smearing
related effects, the tunneling DOS below and above T$_c$ for a SIN
(Superconductor-Insulator-Normal metal) junction
are proportional to $N_{sup}(E)$ and N(E), respectively. Here, N(E), the 
normal
state DOS, is assumed to be constant near E$_F$. We believe that for 
underdoped
high-T$_c$ superconductors N(E) has a depression near E$_F$ which may also
depend on the angle $\theta$ in ab-plane. In this case, the DOS below T$_c$ 
is
given by, $N_{sup}(E,\theta)=N(E,\theta)|E|/\surd(E^2-\Delta(\theta)^2)$. 
Hence
$N(E,\theta)$ can be normalized away from the tunneling spectra at low
temperatures by the tunneling spectra above T$_c$ for ab-plane tunneling at 
a
particular angle $\theta$. For the c-axis tunneling, where the tunneling
conductances below and above T$_c$ are $\int{N_{sup}(E,\theta)}d\theta$ and
$\int{N(E,\theta)}d\theta$, respectively, the normal state DOS cannot be
normalized away from the low temperature spectra. However, the effect of
pseudogap could still be observable in terms of non-conservation of states 
if
there is a depression or hump in N(E).

For an SIS (Superconductor-Insulator-Superconductor) tunneling junction the
tunneling current is given by (neglecting the thermal smearing effects), 
I(V) $ \propto
\int\limits_{0}^{eV}N_{sup}(E,\theta)N_{sup}(E-eV,\theta)dE$ (for T $<$ 
T$_c$)
and  I(V) $\propto \int\limits_{0}^{eV}N(E,\theta)N(E-eV,\theta)dE$ (for T 
$>$
T$_c$). For $|eV| > {\Delta}$, $N_{sup}(E,\theta) \approx N(E,\theta)$ and
hence it turns out that for $|eV|>2{\Delta}$, I(V) and dI/dV are determined
by $N(E,\theta)$ at low temperatures. Thus, the noramalization procedure for
tunneling conductance is still valid for higher bias voltages 
($|eV|>2\Delta$)
for ab-plane SIS tunneling junctions. Furthermore, we know that for a BCS
type SIS tunneling junction the average conductance matches with the
background conductance and so any deviation from this implies some structure
in the normal state DOS.
\begin{figure}[pt]
\centerline{\epsfxsize=1.3in\epsfbox{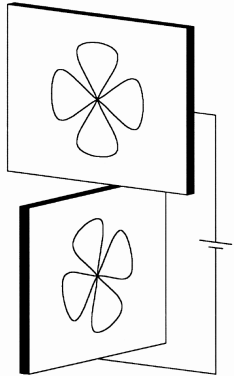}}
\caption {ab-plane tunneling SIS junction geometry. For the SIN junction, 
one of
the superconductors is replaced by a sharp metal foil.}
\end{figure}

In this paper we analyze the ab-plane spectra of underdoped and slightly
overdoped Bi2212 as obtained with SIS and SIN type junctions realized in a 
low
temperature STM. This STM has been used previously for studying the gap
anisotropy\cite{kane-prb-prl} and more recently to observe the
pseudogap\cite{gupta-rapcom} in ab-plane of Bi2212. The crystal growth and
underdoping have been described earlier\cite{gupta-rapcom}. The junction
configuration is as shown in Fig.1. For SIS
tunneling, a single crystal of Bi2212 is cut in the air and the two pieces 
are
attached to two different metal electrodes. The two electrodes are brought
closer, with the two crystals at $90^\circ$ (see fig.1), until a tunneling
current is detected. For SIN tunneling, a
sharpened Pt-Ir foil replaces one of the superconductors. This method has an
advantage in terms of tunneling in a particular direction of the ab-plane as
opposed to c-axis or the break junction configuration. As argued  earlier, 
in
the c-axis tunneling the conductance gives an average DOS for all the angles
in ab-plane. In a break junction, the conductance is also an average over
certain angles in an uncontrolled way. In our junction configuration,
electrons tunnel in a narrow angular cone at a particular angle in the
ab-plane.
\begin{figure}
\centerline{\epsfxsize=2.8in\epsfbox{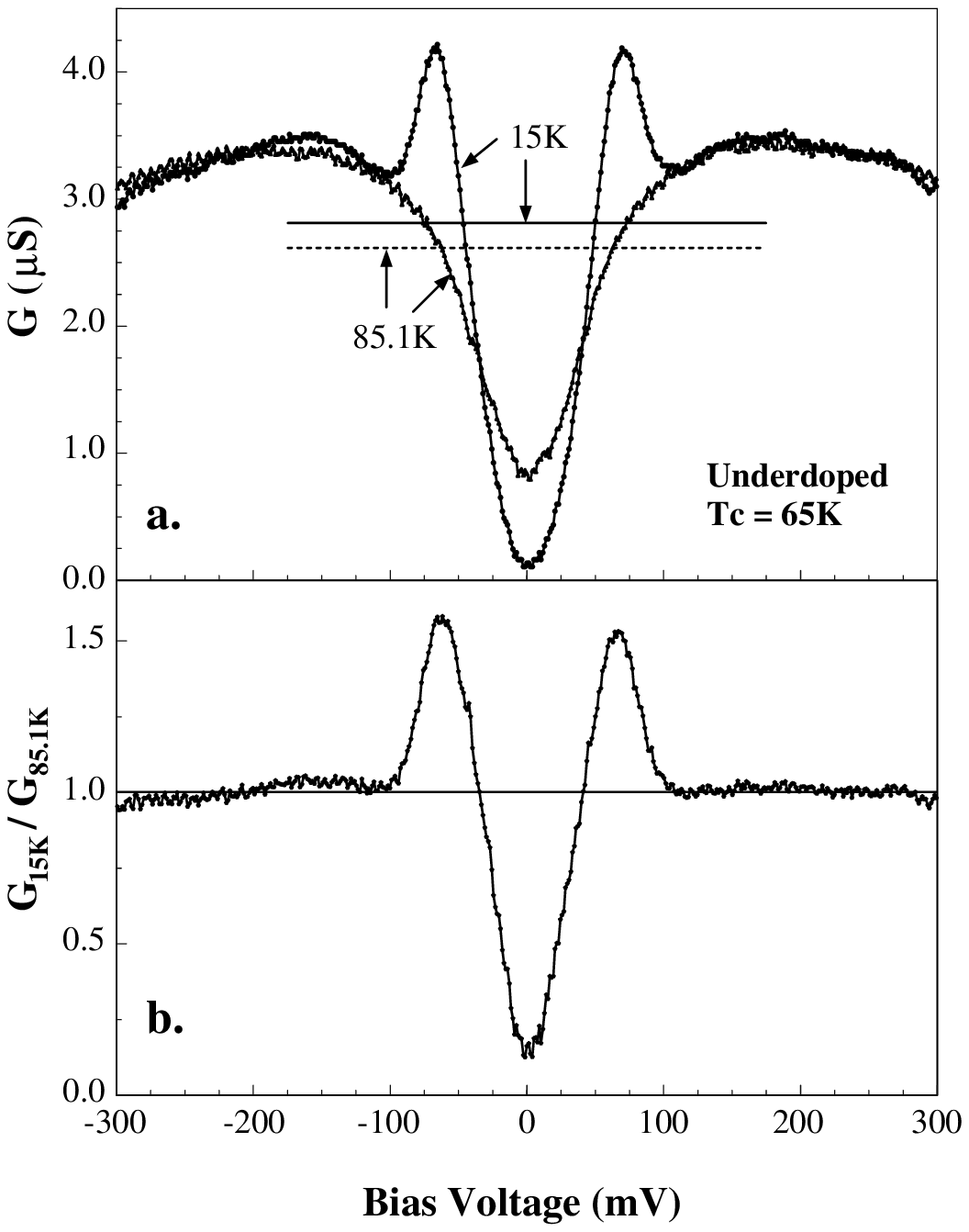}}
\caption{{\bf a.} SIS tunneling conductance spectra for underdoped Bi2212
(T$_c$=65K) at 15K and 85.1K. The spectra are normalized (with no offset) to
equalize the conductances at 300mV. The continuous (dotted) line is the
average conductance between $\pm$175mV for 15K (85.1K) spectrum. {\bf b.} 
The
15K spectrum divided by the 85.1K spectrum. The horizontal line is the
average conductance.}
\end{figure}

The temperature dependent tunneling spectra for underdoped and slightly
overdoped Bi2212 SIN and SIS junctions have been reported
elsewhere\cite{gupta-rapcom,gupta-physica-b}. In Fig.2a we plot a low 
temperature (15K)
and a higher temperature (85.1K) SIS tunneling spectra for a underdoped 
Bi2212
(T$_c$ = 65K) superconductor. The energy gap estimated from the peak-to-peak
separation is about 34mV with the dip feature occuring at about $\pm$ 100mV.
We do not know the exact tunneling direction in the ab-plane and it will not
affect the argument presented here. However, from such a low zero bias
conductance, it is clear that the tunneling angle should be very close to 
the
maximum gap direction. A broad gap-like feature (inside $\pm$ 160mV) is also
clearly visible at low temperatures, together with the superconducting gap.
The gap value and the symmetry of the spectra changes slightly from one
junction to another; however, these three features, peak-dip-hump, are 
clearly
visible in all the SIS spectra for either polarity of the bias voltage. We
believe that the inverted parabolic background (also seen in interlayer
Josephson measurements\cite{josephson}) is a result of the convolution of a
linear background DOS in both the superconductors. It can be shown
analytically that a linear background DOS in the two electrodes of a 
tunneling
junction gives rise to an inverted parabolic background in the tunneling
conductance. Such a linear background has been observed in SIN junctions
previously\cite{pg-renner,miyakawa-john,matsuda}.

One very important point to notice about the spectra is that the 
conservation
of states rule is violated. To analyze this better, an average DOS is
calculated for both the spectra in between the humps ($\pm$ 175 mV). By
average DOS, we mean the integrated conductance in between the specified 
bias
voltages divided by the bias voltage difference. This average conductance is
shown as a continuous (broken) horizontal line in the same figure for 15K
(85.1K). It can be seen from the figure that these lines are well below the
background. This implies that there is a loss of states at low temperature 
as
well as for T $>$ T$_c$. This can be interpreted in terms of an underlying
structure in the DOS other than the qusiparticle type gap. The 
non-conservation feature is
consistently reproducible for different junctions with junction resistance 
varying between 10k$\Omega$ to 10M$\Omega$. This rules out the possibility of 
capacitive effects giving rise to a depression near zero bias.

From Fig.2a the average DOS above T$_c$ seems to be lower than that below
T$_c$, which is quite contrary to the belief that the pseudogap gets weaker
with increasing temperature. We do not quite understand this fact and it
does not change the argument in terms of the non-conservation of states for
the same spectrum; however, we want to point out that the two spectra have
been normalized so as to match the conductances at 100mV. This affects the
relative position of the average DOS of the two different spectra. This
normalization procedure to superimpose the two spectra for comparison 
purposes
has been used for all the spectra described in this paper. 

Further, we divide the 15K curve by 85.1 K curve to normalize away the
background DOS. This is plotted in Fig. 2b. After this normalization, the
dip-hump structure disappears and the average DOS matches with the 
background.
As pointed out earlier, this kind of normalization procedure is valid only 
for
the biases outside the gap structure and in terms of demonstrating the loss 
of
states. This normalization procedure was used in the pioneering work of
McMilan\cite{parks-book} on strong coupling superconductors for removing the
normal state background from the SIS and SIN junctions. This procedure was
found to be necessary in terms of removing the normal state background 
effects
to deduce the strong coupling features from the tunneling conductance. 
DeWilde
et. al.\cite{dewilde-john} modeled the background with a sloped straight 
line;
however, no such normalization was carried out with the experimentally 
measured
tunneling conductance above T$_c$ to remove the normal state background DOS.

Several other groups have clearly seen such non-conservation features with
underdoping in the tunneling DOS in c-axis tunneling junction
configuration\cite{loss-of-states} while DOS for the overdoped Bi2212,
where there is no pseudogap, follows the conservation of
states rule\cite{renner-prb}, consistent with our interpretation. Matsuda
et. al.\cite{matsuda} considered
the pseudogap as a band structure effect, however, no quantitative analysis 
was
done to determine the loss in states. Although they normalized the low
temperature spectra with the 100K spectra, the spectra looked very 
unrealistic
after normalization. Since the c-axis spectrum is an anglular averaged
ab-plane spectrum as we have discussed previously in this paper, such
normalization procedure is not valid for the c-axis tunneling spectra given 
the highly
anisotropic nature of the two gaps.
\begin{figure}
\centerline{\epsfxsize=2.8in\epsfbox{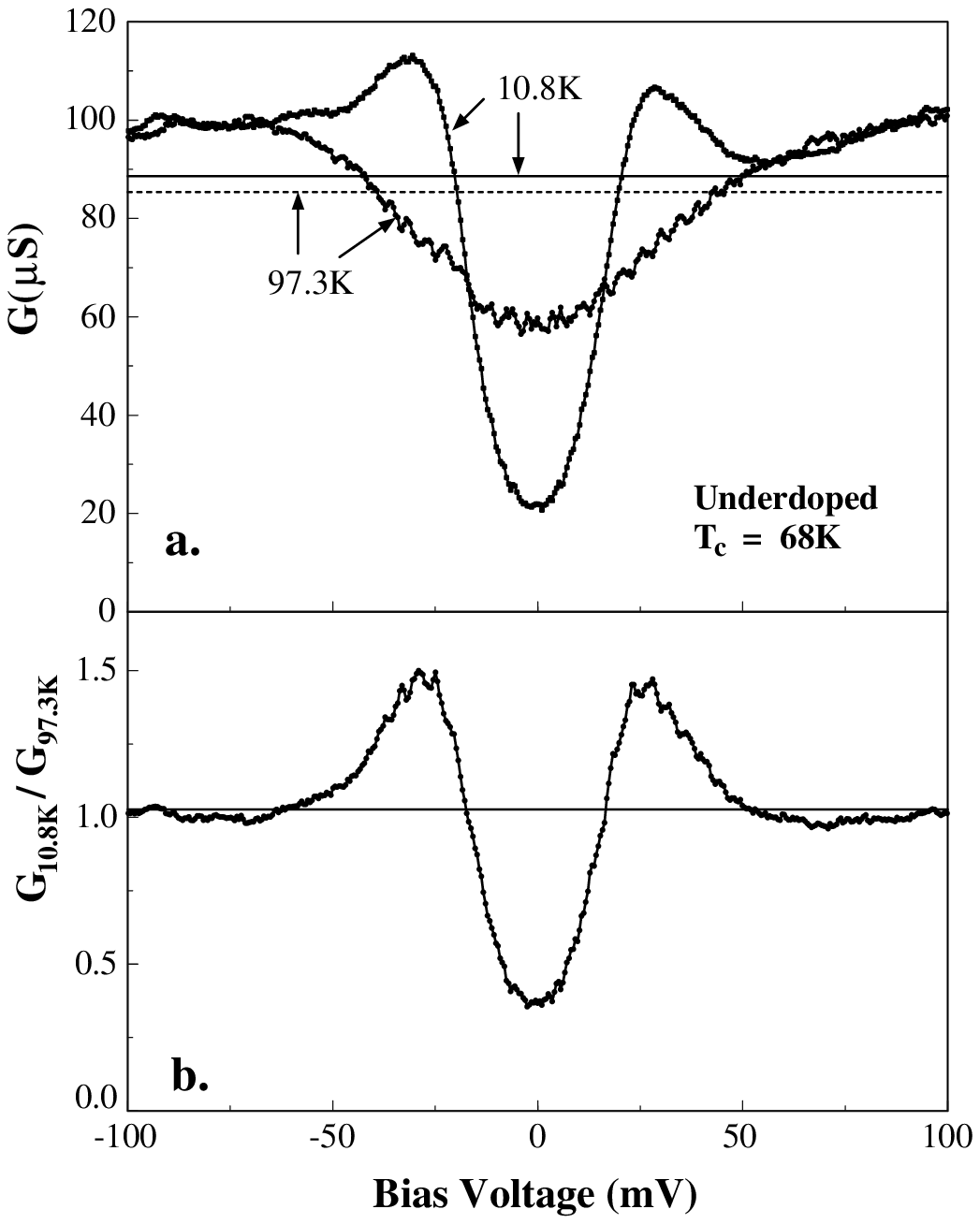}}
\caption{{\bf a.} SIN tunneling conductance spectra for underdoped Bi2212
(T$_c$=68K) at 10.8K and 97.3K (normalized at 100mV without offset). The
continuous (dotted) line is the average conductance between $\pm$100mV for
10.8K (97.3K) spectrum.{\bf b.} The 10.8K spectrum divided by 97.3K 
spectrum.
The horizontal line is the average conductance between $\pm$100 mV.}
\end{figure}

The tunneling spectra for a SIN junction for underdoped Bi2212
(T$_c$ = 68K) are plotted in Fig.3a for 10.8K and 97.3K. The SIN spectra 
have
less sharp and more asymmetric features as compared to the SIS spectra. The
spectra have been normalized so that the conductances at 100mV are the same.
The dip structure seen on the positive bias side is actually consistent with
the dip seen by other groups on the negative side. The bias voltage in this
case is applied to the metal electrode while the sample is at the zero
potential. The dashed (continuous) horizontal line shown in the figure is 
the
average DOS for 10.8K (97.3K), which is well below the background 
conductance
for both the temperatures implying the loss of states. One very interesting
feature about these and the SIS spectra (Fig.2a) is that the backgrounds 
below
T$_c$ and above T$_c$ match remarkably well and the dip feature seen on the
right for T $<$ T$_c$ seems to be a result of the product of the background
depression in DOS and a BCS like gap DOS. This is a very strong evidence in
support of the fact that the same pseudogap exists for both temperatures,
above and below T$_c$.

	In Fig.3b, we divide the low temperature curve by the high temperature
one. The average tunneling conductance of the normalized spectrum matches 
well
with the background implying that the conservation of states is recovered.
Moreover, the normalized tunneling spectrum is symmetric and most of the dip
feature on the right is absent. A little depression, which is still present 
as
the dip feature, can be attributed to a slight weakening of the pseudogap 
with
increasing temperature.
\begin{figure}
\centerline{\epsfxsize=2.8in\epsfbox{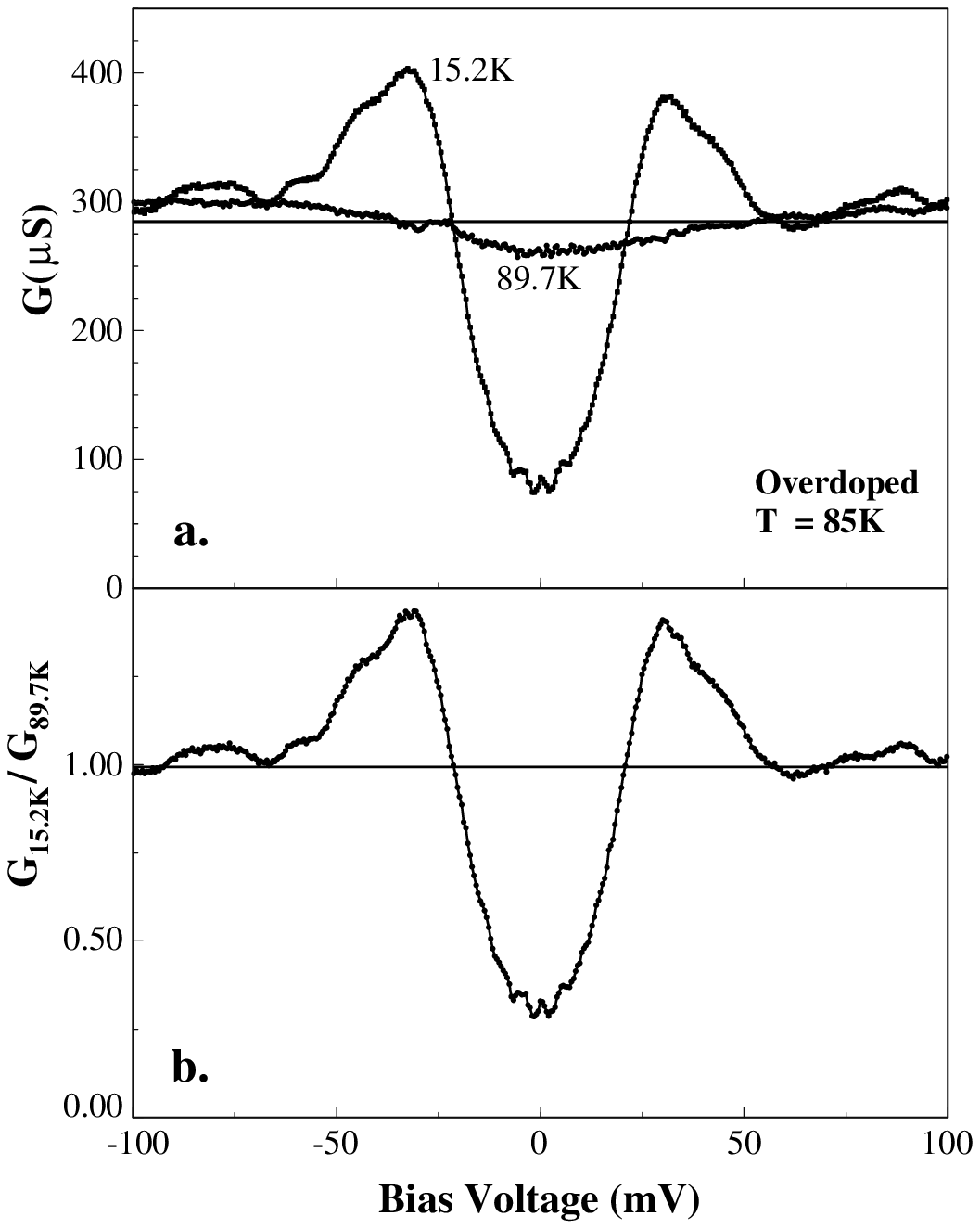}}
\caption{{\bf a.} SIN tnneling conductance spectra for slightly overdoped
Bi2212 (T$_c$=85K) at 15.2K and 89.7K (normalized at 100mV without offset). 
The
horizontal line is the average conductance which, in this case, is equal for
both the spectra. {\bf b.} The 15.2K spectrum divided by 89.7K spectrum with
the horizontal line as average conductance between $\pm$100 mV.}
\end{figure}

To compare with the overdoped case, we plot the tunneling spectra for a SIN
junction at 15.2K and 89.7K for slightly overdoped Bi2212 (T$_c$ = 85K). 
From
the 15.2K spectrum the average conductance was calculated and we find that
the loss of states is negligible. This means that the pseudogap in this
compound is much weaker than the underdoped one. The low temperature 
spectrum has
an asymmetry, smaller than the underdoped SIN spectrum, and a dip feature 
for
the positive bias. This asymmetry is also present in 89.7K spectrum. On
normalization of the 15.2 K spectrum with the 89.7K spectrum the asymmetry
disappears, however, the dip feature still persists.

Recent c-axis tunneling measurements from two different 
groups\cite{pan-kapitulnik-inhom}
found that the surface of Bi2212 is quite inhomogeneous with some areas 
having much
larger gap compared to the superconducting gap. It could be argued that
the non-conservation features we observed could be a result of the 
contribution from
such large gap regions which may not conserve states, given that the 
junction area in
our tunneling configuration is larger than a STM tip. In this case, it 
should be
possible to subtract out the contribution from such regions by subtracting 
the normal state spectra from the low temperature spectra. We carried out this 
subtraction and found that the state conservation was
almost recovered.This can also be seen from Fig. 2a,3a, and 4a, since the 
average conductances for the two spectra are almost equal. It should be 
noted that this procedure is
sensitive to the relative normalization factor between the two spectra as 
mentioned earlier.
However, the subtraction procedure does not remove the dip features from the 
SIS and SIN
spectra and also it does not symmetrize the SIN spectra completely. Although 
we cannot
rule out this inhomogeniety scenario completely, it does not affect our 
argument in terms of
non-conservation of states for the underdoped Bi2212 and hence the 
coexistence of the two gaps.

To summarize, from our ab-plane tunneling studies on Bi2212, we find that 
the
states conservation rule is violated for the underdoped material. By 
normalizing the low
temperature spectra with the T $>$ T$_c$ spectra, the conservation of
states rule is recovered. Moreover, the dip-hump feature which is observed
below T$_c$ disappears on normalization. From this, we conclude that the
pseudogap does not evolve into the superconducting gap at T$_c$; rather, the
two gaps coexist below T$_c$. Furthermore, the dip-hump feature can be
interpreted in terms of the coexistence of the two gaps, one being slightly
larger than the other. This coexistence of two gaps can rule out the 
scenario
that the superconducting gap and pseudogap arise from the same origin, i.e.
pair formation with and without coherence, respectively.

We would like to acknowledge a brief discussion we had with A. J. Millis on
non-conservation of states. This work was supported by NSF grant DMR997201.

%%%%%%%%%% REFERENCES %%%%%%%%%%


\begin{references}
\bibitem{loss-of-states}
For eg., see fig.1 of ref.7, fig.3 of ref.8; the tunneling spectra at low
temperatures show the loss of states for the underdoped
superconductors. This systematic loss of states can be seen from the 
decrease in the
coherence peak height and enhanced dip structure with underdoping. This loss 
of states
gets weaker with overdoping.
\end{references}
\end{document}